\documentclass[useAMS,usenatbib]{mn2e}

\usepackage{graphicx}
\usepackage{lscape}

\usepackage{biblio}

\usepackage[usenames]{color}
\usepackage[normalem]{ulem}
\newcommand{\rk}[2]{#2}

\newcommand{\msun}{M_{\odot}}
\newcommand{\oii}{[\mbox{O\,{\sc ii}}]}
\newcommand{\ewhd}{\rm EW(H\delta)}

\title[E+A post-starburst SEDs]{
  The role of E+A and post-starburst galaxies \\
  II. Spectral energy distributions and comparison with observations}
\author[Falkenberg, Kotulla \& Fritze]{M. A. Falkenberg$^{1}$\thanks{E-mail:
    atyra@astro.physik.uni-goettingen.de}, R. Kotulla$^{2}$\thanks{E-mail:
    r.kotulla@herts.ac.uk, u.fritze@herts.ac.uk} and U. Fritze$^{2}$\\
  $^{1}$Institut f\"ur Astrophysik, Georg-August
  Universit\"at G\"ottingen, Friedrich-Hund-Platz 1, 37077 G\"ottingen, Germany\\
  $^{2}$Centre for Astrophysics Research, University of Hertfordshire, College
  Lane, Hatfield AL10 9AB, United Kingdom}
\begin{document}

\date{Accepted 2009 May 5.  Received 2009 March 20; in original form 2008 June 12}

\pagerange{\pageref{firstpage}--\pageref{lastpage}} \pubyear{2009}

\maketitle

\defcitealias{FalkenbergFritze09}{Paper~I}

\label{firstpage}

\begin{abstract}
In a previous paper (Falkenberg, Kotulla \& Fritze 2009, astro-ph/0901.1665) we
have shown that the classical definition of E+A galaxies excludes a significant
number of post-starburst galaxies. We suggested that analysing broad-band
spectral energy distributions (SEDs) is a more comprehensive method to select
and distinguish post-starburst galaxies than the classical definition of
measuring equivalent widths of (H$\delta$) and {\oii} lines.

In this paper we will carefully investigate this new method and evaluate it by
comparing our model grid of post-starburst galaxies to observed E+A galaxies
from the MORPHS catalog.

In a first part we investigate the UV/U-optical-NIR SEDs of a large variety in
terms of progenitor galaxies, burst strengths and timescales of post-starburst
models and compare them to undisturbed spiral, S0 and E galaxies as well as to
galaxies in their starburst phase. In a second part we compare our
post-starburst models with the observed E+A galaxies in terms of Lick indices,
luminosities and colours. We then use the new method of comparing the model
SEDs with SEDs of the observed E+A galaxies.

We find that the post-starburst models can be distinguished from undisturbed
spiral, S0 and E galaxies and galaxies in their starburst phase on the basis of
their SEDs. It is even possible to distinguish most of the different
post-starburst by their SEDs. From the comparison with observations we find that
all observed E+A galaxies from the MORPHS catalog can be matched by our
models. However only models with short decline timescales for the star formation
rate are possible scenarios for the observed E+A galaxies in agreement with our
results from the first paper (see Falkenberg, Kotulla \& Fritze 2009a).

\end{abstract}


\section{Introduction}
The morphology-density relation \citep{Oemler1974ApJ, Dressler80} and
the Butcher-Oemler effect \citep{Butcher1978ApJ,Butcher1984,Yang2004}
indicate a transformation from late- to early-type galaxies during the lifetime
of a galaxy in high density environments. Over the past a wide range of
transformation processes have been proposed, acting in different environmental
densities and on different timescales: Important mechanisms are mergers,
harassment, gas stripping and strangulation.

Merging of two or more galaxies is an interaction most efficient at low relative
velocities. If at least one gas-rich galaxy is involved in the merger this
triggers a starburst and a subsequent morphological transformation, in general
into an early-type galaxy \citep[e.g.][]{BarnesHernquist96}. Galaxy harassment
\citep{Moore+96,Moore+98} are multiple high-speed encounters in the densest
regions of the cluster, but can have effects out to large cluster radii
\citep{Moran+07}. This harassment process can also be connected to a starburst
\citep{Lake+98}. Finally there are gas stripping and strangulation, two related
processes in which the infall into the galaxy cluster and contact to the hot
X-ray gas within the cluster leads to a truncation or termination of star
formation (SF). One example is ram pressure stripping (RPS) of the infalling
galaxy's gas. The difference between truncation and termination is the timescale
on which the gas is removed: In the case of truncation the RPS is strong enough
to strip the gas out of the disk, truncating SF on timescales of $\approx$ 100
Myr. Termination on the other side primarily strips off the halo gas, leading to
a halting of SF on longer timescales of $\approx$ 1 Gyr \citep{KawataMulchaey08,
  Moran+06, Barr+07, KodamaBower01}.

To evaluate the possible mechanisms that take place during this transformation
in cluster environments like harassment, gas stripping, strangulation and
mergers \citep[see, e.g.,][]{Poggianti2004PoS}, the investigation of
post-starburst galaxies as intermediate state between late- and early-type
galaxies is necessary. Several earlier studies \citep[e.g.,][]{Barger+96,
  Leonardi+96} found that a significant fraction of cluster members show
spectral features characteristic for a recent starburst in the last 2 Gyr, and
that those starbursts affected large fractions of the total galaxy mass. If
those starbursts are triggered by merging this can explain the transformation of
disk-dominated into bulge-dominated (S0) galaxies. This would naturally explain
the resulting decrease in the fraction of blue galaxies in intermediate redshift
clusters as compared to local samples \citep{Dressler80, Dressler1997,
  Poggianti+08}. An alternative explanation for the same effect is the
truncation of SF in spirals and the subsequent passive evolution, leading to a
fading and migration onto the red sequence \citep{Treu+03, DeLucia+04}.

\citet[Table 6]{Dressler1999} introduced a classification scheme based on the
equivalent widths of {\oii} and H$\delta$ lines. The relevant classes for the
present paper are k, k+a, a+k and e(a). e(a) galaxies have spectra that have
strong Balmer absorption $\rm \ewhd \geq 8\,\AA$ but {\oii} in emission. The
remaining classes a+k, k+a, and k all have no detecable {\oii} emission, but
differ in the strength of their balmer absorption, with a+k having strong
($\ewhd\ge8\,\AA$), k+a moderate ($3\,\AA\le\ewhd\le\,8\AA$) and k-type spectra
weak ($\ewhd\le3\,AA$) H$\delta$ features.
Another common term for post-starburst galaxies is the class of E+A
galaxies. E+A galaxies are described by a superposition of an elliptical galaxy
spectrum (hence the E) with the spectra of A-type stars with strong Balmer
absorption lines (the A in E+A), making if a more general class including a+k
and k+a galaxies. In the remainder of this paper we will therefore use the term
E+A interchangeably with the combination of k+a/a+k galaxies. Note that we
assign positive signs to $\ewhd$ if H$\delta$ is in absorption, while positive
signs mean emission of the $\oii$ lines.

In a first paper \citep[\citetalias{FalkenbergFritze09}
  hereafter]{FalkenbergFritze09} we focused on modelling E+A galaxies
\citep{Goto2004AA, Poggianti1999ApJ} with our galaxy evolution code
GALEV. For this purpose we imposed different combinations of a starburst and
halting of star formation (SF) with different timescales and beginning times on
spiral galaxy models.

We found that the classical definition of E+A galaxies by measuring the
equivalent width of Balmer absorption and {\oii} emission lines
\citep{Goto2004AA,Poggianti2004, Yang2004} excludes a significant number
of post-starburst galaxies, in particular those in stages with still some SF
going on after a burst. By investigating model spectra we predict that
Spectral Energy Distributions (SEDs) extending from UV/U through optical and
eventually near-infrared (NIR) passbands 1.) allow a better distinction of E+A
progenitors and scenarios and 2.) give a comprehensive census of all
post-starburst galaxies of which, as we have shown, the E+As are only a
subclass.

In the first part of this paper we investigate the full SEDs of our models.
The second part shows a comparison of models with observations from the
MORPHS group. 

\section{GALEV models}
\label{sect:galev}
To model the galaxy transformation processes we use our GALEV evolutionary
synthesis models \citep{Bicker2004A&A,Anders2003A&A,Kotulla+09}. The full
description of the used models and the modifications to compute the {\oii}
emission line strength and $\rm H\delta$ equivalent widths can be found in
\citetalias{FalkenbergFritze09}, hence we only give a short overview here. GALEV
models allow to trace the spectral and photometric evolution of galaxies with
arbitrary star formation histories (SFHs). We use stellar evolution data from
the Padova group \citep{Bertelli+94} including the important TP-AGB phase. To
accurately model gaseous emission lines we supplement these data with data for
the Zero Age Main Sequence \citep[ZAMS, see][for details]{Anders2003A&A}. We
further use the stellar template library of \cite{Lejeune+97,
  Lejeune+98}. Convolving the isochrones with a \cite{Salpeter55}-IMF with
mass-limits of $0.1$ and $100$ M$_\odot$ and stellar template spectra yields
isochrone spectra that can then be weighted with the SFH of each galaxy type and
integrated to yield the galaxy spectrum. Magnitudes in the various bands are
then computed by convolving the galaxy spectra with the filter response curves
and applying the appropriate zero points.

\subsection{Undisturbed galaxies}
SFHs of undisturbed spectral types are calibrated using a wealth of
observational data such as integrated spectra and colours, star formation rates
(SFRs), gas-fractions, gaseous and stellar metallicities, and mass-to-light
ratios (see \cite{Bicker2004A&A} and \cite{Kotulla+09} for a detailed
description of this calibration). Note that our galaxy types E/S0 and Sa to Sd
are meant to represent spectral types resulting from the chosen SFH. In the
local universe those spectral types correlate very well with morphological
types, but this might not hold to arbitrarily high redshifts. To fully reproduce
the observed properties of local, \rk{galaxy templates}{observed galaxies} it is
furthermore necessary to match the observed absolute magnitudes for the average
representative of each galaxy type. Since colours and spectra only depend on the
SFH (which is assumed to be independent of galaxy mass) we chose a total galaxy
mass (including both stars and gas) that matches after a Hubble time the average
absolute magnitudes of galaxies from the Virgo galaxy cluster.

\subsection{Modelling galaxy transformation scenarios}
To model galaxy transformation scenarios we add a starburst followed be a
truncation of SFR or a pure truncation of SF depending on the transformation
process (see Sect. 2.4 of \citetalias{FalkenbergFritze09} for
details). Starbursts are described by three parameters: 1) the time $\rm
t_{burst}$ or redshift $\rm z_{burst}$ of the burst; 2) the burst strength $b$,
i.e. the fraction of gas available at the onset of the burst that is converted
into stars during the burst; 3) the decline timescale $\tau$ of the exponential
declining SFR after the burst. The full parameter set used in the rest of this
paper are summarized in Table \ref{tab:paramgrid}. Note that this model grid
does not cover the full parameter grid. The reason for that is that we do not
list any models that do not fulfill the E+A criterion (e.g. models with Sa
progenitors and/or weak bursts). In this work we identify galaxies as k+a type
galaxies if they have H$\delta$ equivalent widths $\rm EW(H\delta)\ge 5\,\AA$
and only weak {\oii} emission lines with $\rm EW(\oii)<5\,\AA$.

\begin{table}
\caption{Parameter space covered by our grid. The upper part lists the models we
  use to demonstrate the effects of different starbursts scenarios on the
  observed SEDs, the lower part gives details on the models we compare to
  observations (Sect. \ref{chapter_obs}). The latter also include evolutionary
  and cosmological corrections.}
\centering
\begin{tabular}{llll}
Progenitor & $\rm t_{burst}$ [Gyr] & b [\%] & $\tau$ [Gyr] \\
\hline
Sa & 6  & 50 & 0.1 \\
   & 11 & 30 & 1.0 \\
\hline
Sd & 6  & 50 / 70 & 0.1 \\
\hline
\\
\\
Progenitor & $\rm z_{burst}$ & b [\%] & $\tau$ [Gyr] \\
\hline
Sc & 0.44 & 70 & 0.1, 0.3 \\
   & 0.57 & 70 & 0.1, 0.3 \\
\hline
Sd & 0.44 & 0  & 0.1 \\
   & 0.57 & 0  & 0.1 \\
   &      & 70 & 0.1, 0.3 \\
   & 0.90 & 70 & 0.1 \\
\hline
\end{tabular}
\label{tab:paramgrid}
\end{table}

\subsection{Impact of dust extinction}
Besides the SFHs the amount of dust extinction is the dominating factor
influencing the colours of galaxies.  However, in the present study we are not
dealing with galaxies with ongoing starbursts, but rather galaxies in their
later, post-starburst phases with no or only little remaining SF. In
\citetalias{FalkenbergFritze09} we found that only galaxies with strong bursts
$b\simeq70\%$ reach the H$\delta$ strong phase. By the time it reaches this
stage the galaxy should be essentially gas-free, since most of the gas was
consumed in the past burst and the remaining gas ejected due to feedback
processes (see Sect. \ref{sect:gasfate}). This is also confirmed by
non-detection of the majority of k+a galaxies in HI observations
\citep{Goto2004AA, MillerOwen01}. Since the amount of dust is linked to the gas
reservoir we do not expect significant amounts of dust in those late
post-starburst phases. This is also supported by \cite{Balogh+05} who used dust
sensitive optical-NIR colours and found that their spectra are not significantly
affected by dust. Note that this does not contradict the findings of (very)
dusty \textbf{ongoing} starbursts during earlier phases of galaxy
transformation, e.g. the e(a) phases \citep{Shioya+01, Balogh+05, Dressler+08,
  Poggianti+08}.


\section{Spectral Energy Distributions of Post-Starburst
  Galaxies}\label{chapter_sed}

With SEDs it is possible to investigate the changes in magnitude for each
filter as well as in colours of a galaxy at the same time. This enables an
  easy comparison of SEDs of undisturbed galaxies with SEDs of galaxies
  undergoing a starburst or SF truncation event.

\begin{figure}
\includegraphics[width=\columnwidth]{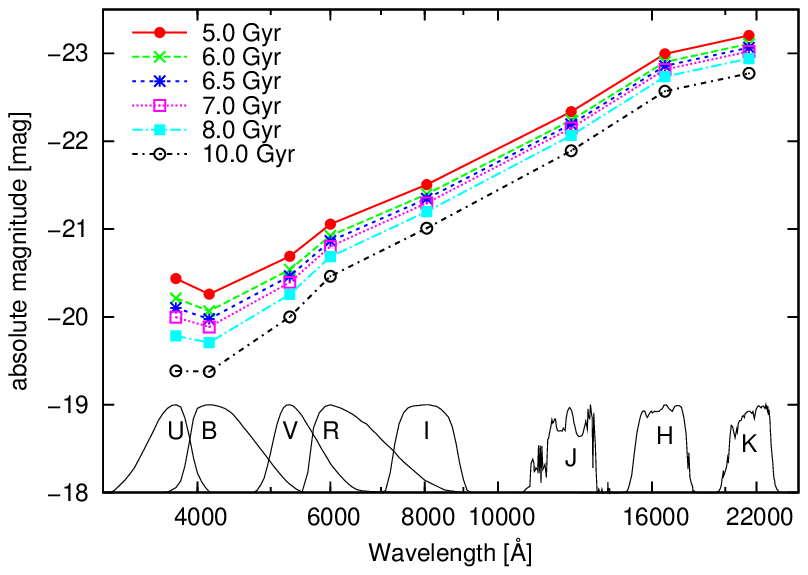}
\includegraphics[width=\columnwidth]{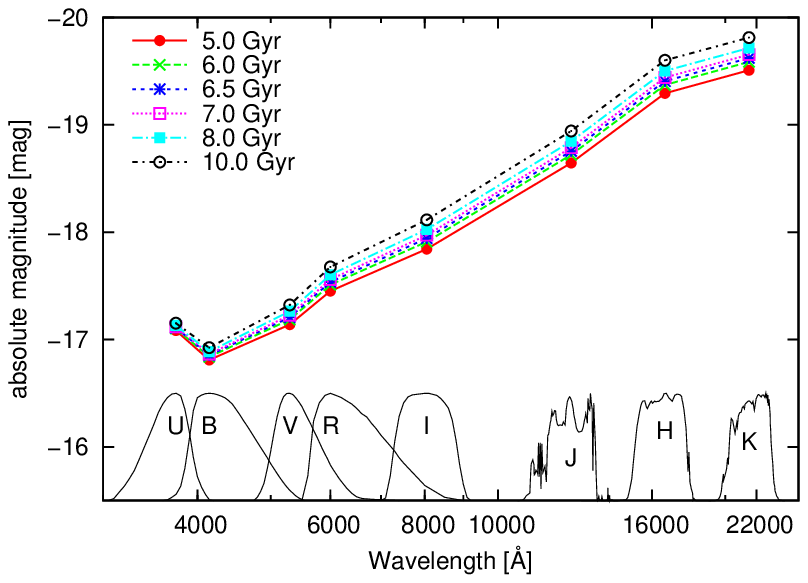}
\caption{SEDs for an undisturbed Sa (top panel) and Sd model (bottom panel) with
  ages between 5 and 10 Gyr. The connecting lines are only meant to guide the
  eye.}
\label{sed_undist_Sa}
\end{figure}

In Fig. \ref{sed_undist_Sa} we show the SEDs of undisturbed Sa and Sd
models. The symbols show magnitudes in the Vega system for the Johnson U to
  I and Bessell-Brett \citep{Bessel1988PASP} J to K filters, put at their
  respective central wavelengths. The lines connecting the symbols are only
meant to guide the eye.  We recall that the luminosities of our undisturbed
model galaxies at an age of 13 Gyr are gauged to agree with average observed
B-band luminosities of the respective galaxy types Sa, Sb, Sc and Sd as observed
in the Virgo cluster. However, real galaxies show a spread in absolute
  magnitudes. The observed 1 $\sigma$ ranges for galaxies of each type,
  derived from luminosity functions of galaxies in the Virgo cluster
  \citep{Sandage1985AJa, Sandage1985AJb}, are given in Table \ref{sigma}.

\begin{table}
\caption{1 $\sigma$ ranges for observed galaxies in the Virgo cluster, taken
  from \citet[Fig. 10]{Sandage1985AJa} and
  \citet[Fig. 4]{Sandage1985AJb}}\label{sigma} \centering
\begin{tabular}{l l }\hline \hline
Morph. Type & 1 $\sigma$ range [mag]\\
            & apparent magnitudes   \\ \hline
E / S0      & 1.5 \\
Sa          & 0.9 \\ 
Sb          & 1.1 \\
Sc          & 0.9 \\
Sd          & 0.8 \\ \hline
\end{tabular}
\label{tab:sigmaranges}
\end{table}

\subsection{Luminosity evolution}
Galaxies become fainter and bluer towards later types. Magnitudes of the typical
Sa-type galaxy (see top panel of Fig. \ref{sed_undist_Sa}) hence range from
$\rm\approx -19\,mag$ in the U-band to $\rm\approx -23\,mag$ in the K-band. The
typical Sd-galaxy in comparison (bottom panel of Fig.  \ref{sed_undist_Sa}) is
fainter by $\rm\approx2\,mag$ in the U-band and $\rm\approx 3.5\,mag$ in K.

The time evolution of typical brightnesses differs between galaxy types.
Early-type spirals, such as Sa-type galaxies, with their declining star
formation rates (SFRs) are bright early in their life and become fainter with
time. This effect is most dramatic in the U-band because short wavelengths are
most affected by the lower SFR at later times, while long wavelengths show
smaller changes.  For Sd galaxies and their constant SFR scenario the blue part
of the spectrum, which is dominated by massive, hot stars with short lifetimes,
very quickly reaches an equilibrium state and remains constant with
time. However, cooler, low-mass and hence long-living stars slowly accumulate
with time, leading to an increase of luminosity with time in the long-wavelength
bands.

Undisturbed Sb and Sc models evolve similar to the Sa model, while the
luminosity difference between the different ages is smaller, i.e. the SEDs for
different ages lie closer together. Luminosities of the Sb and Sc
models vary between those of the Sa and Sd models.

Note the change with time in the shape of the SED at the short wavelength end
in case of the Sa model with its strongly decreasing star formation rate as
opposed to the case of the constant SFR Sd model.

\begin{figure}
\includegraphics[width=\columnwidth]{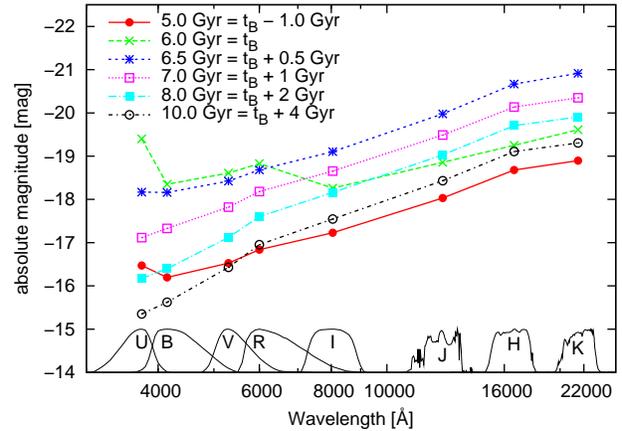}
\caption{SED of an Sd burst model with ages between 5 and 10 Gyr (Sd, burst of
  70\% at a galaxy age of 6 Gyr and decline time $\tau=0.1$ Gyr).}
\label{sed_burst_max}
\end{figure}

Fig. \ref{sed_burst_max} shows the SEDs of an Sd model undergoing a burst
leading to the maximum peak value for EW(H$\delta$) and the longest E+A phase
of our model grid \citepalias[see][]{FalkenbergFritze09}. The burst starts at a galaxy age of
$\rm t_{burst}=6$ Gyr and is described by a sudden increase of the SFR to a
peak value and a subsequent exponential decline of SFR with time on a
characteristic timescale $\tau$, here $\rm\tau=10^8\,yr$. The burst strength,
in this case 70\%, is defined as the fraction of gas available at the onset of
the burst and converted into stars during the burst.

At 5 Gyr the model is still undisturbed, the SED identical with the SED at 5 Gyr
in the bottom panel of Fig. \ref{sed_undist_Sa}.  At 6 Gyr all luminosities of
the burst SED have increased and the galaxy is now much brighter than the
undisturbed model at any time. The overall shape of the SED has changed,
i.e. the starburst galaxy model is even bluer than the undisturbed Sd-model. U-B
colours, which appear as kink between U and B in SEDs like the one in
Fig. \ref{sed_burst_max}, can reach $\rm -1\,mag$ for the strongest bursts. This
kink is, although less prominent, also present in SEDs of undisturbed spiral
galaxies, but absent in SEDs of nearby S0 and E galaxies.

At an age of $6.5$ Gyr, $0.5$ Gyrs after the burst, the galaxy already is in its
H$\delta$-strong phase, but has still measurable {\oii} lines.  We recall
  that we consider a threshold of EW(\oii)$=\rm 5 \AA$; only galaxies with a
  lower equivalent width are regarded as E+A galaxy with absent {\oii}
  lines. This threshold of 5 \AA\ is the upper threshold for EW(\oii) from
  \citet{Dressler1999} \citepalias[see also][]{FalkenbergFritze09}. Since the
beginning of the burst luminosities in I to K have increased, while the U to R
luminosities already started to decrease, but are still brighter than the
undisturbed model before the burst. The H$\delta$-strong model is now redder
than the starburst and undisturbed Sd models. 
the SED of the undisturbed Sd model,

The SED 1 Gyr after the burst shows the typical SED of an E+A galaxy, for
which the EW(\oii) is now lower than 5 \AA.  For the SEDs at
7, 8, and 10 Gyr, all luminosities decrease with increasing age. The shape of
the SED evolves with time into the SED of an early-type galaxy, which is even
redder than a galaxy in the H$\delta$-strong and E+A phase.

\subsection{Impact of burst parameters on the luminosity evolution}
In general the increase in luminosities and change in colours during the burst
and the entire post-starburst phase, i.e. the phase in which the SEDs can be
distinguished from those of the undisturbed spiral galaxy on the one hand, and
from the S0/E galaxy SEDs on the other, is influenced by the impact of the
burst. This impact in turn depends on burst strength b, decline time $\tau$,
onset of the burst, and galaxy type.

Progenitor galaxies of later type show more pronounced differences between the
undisturbed, blue model and the post-starburst phase of the burst
model. Spectral transformation into S0-type remnants is also slower for those
galaxies since they overall have a relatively young stellar population which
has to age first.

Bursts with earlier onset times produce more luminous starburst and
post-starbursts phases. The same holds true for increasing burst strength. The
reason for both effects is our definition of the burst strength. The amount of
stars formed during the burst is given by the available gas-mass at the
onset of the burst multiplied with the burst strength. From there it follows
naturally that earlier onset times correspond to a larger available gas-mass
to be turned into stars as does a larger burst strength.

Halting of star formation on the opposite leaves brighter galaxy remnants if
it happened at a late stage, because until then star formation in the
undisturbed galaxies leads to a continuous built-up of stellar mass.

Bursts with longer decline time evolve slower in both colour and luminosity
towards their postburst-phases and finally into S0-type galaxies. Also the
luminosity increase at the peak of the burst decreases with longer decline
times of the burst.

In the case of a slow decline of SFR with $\tau=1$ Gyr the evolution takes
$\approx 3.5$ Gyr more time than in the case of a short decline with
$\tau=0.1$ Gyr. As mentioned above also the increase in luminosity is less
prominent for the first case and colours do not reach as blue colours as in
the latter case.

\subsection{Impact of bursts on the colour evolution}
During the burst and post-burst phases the U-B colour experiences the largest
changes. Colours in the red part of the optical, e.g. V-I or R-J, also change
slightly, while H-K colours hardly change at all. This can be explained by the
stellar populations dominating each of those regions of the spectrum. The U and
to a lesser degree the B to R-bands are dominated by young, hot stars which
evolve rapidly with the declining star formation. NIR luminosities on the
opposite are dominated by cooler and longer-lived stars and hence the increase
in luminosity in these bands is roughly proportional to the increase in stellar
mass, leaving the resulting colours essentially unchanged.

The shape of the long wavelengths of the SED from I to K-band hence looks
similar in all post-starburst galaxies, while galaxies can have very different
luminosities from -23.5 to -19 mag in the K-band. In general, all SEDs become
fainter and redder after the burst with increasing age. For weak bursts
galaxies can even become fainter than their undisturbed progenitor, since we
assume that star formation declines exponentially to zero after the burst.

In the case of SF truncation or termination the SED immediately becomes
fainter than that of the undisturbed spiral galaxy, so there is no initial
increase in luminosity characteristic for the galaxies with starburst.  While
the U to R part of the SED very quickly adopts the shape of an S0 galaxy SED,
the I to K part of the SED remains very similar to that of the undisturbed
spiral galaxy.

\subsection{Progenitors and remnants}
\begin{figure}
  \includegraphics[width=\columnwidth]{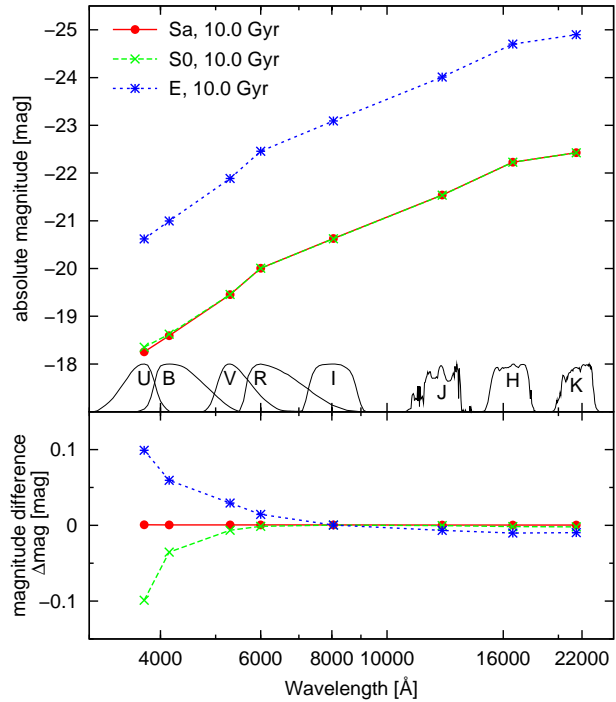}
  \caption{SED of an Sa burst model (50\% burst at 6 Gyr, decline time
    $\tau=0.1$ Gyr) at 10 Gyr compared to the \rk{S0 and E
    template}{undisturbed E and S0 models}.}
  \label{sed_S0_Sa_10}
\end{figure}

In the upper part of Fig. \ref{sed_S0_Sa_10}, we show an Sa-type model with a
burst of 50\% at 6 Gyr and a decline time of 0.1 Gyr with an age of 10 Gyr.  We
also show for comparison the SED \rk{templates}{} of an S0 and elliptical
\rk{template}{model}, both normalized to match the typical B-band luminosities
for their particular type as taken from \citet[Fig. 10]{Sandage1985AJa} and
\citet[Fig. 4]{Sandage1985AJb}.

The SED of the 10 Gyr old Sa post-burst model is essentially indistinguishable
from that of the S0 galaxy. The luminosity difference between the SED for the E
galaxy and the model SED is slightly larger than the 1 $\sigma$ luminosity
ranges of S0 and elliptical galaxies of 1.5 mag as given by \citet[see also
    Table \ref{tab:sigmaranges}]{Sandage1985AJa}. The Sb and Sc model
luminosities match the luminosities of \rk{the S0 template}{observed S0
  galaxies} within the 1 $\sigma$ range, but cannot reach the luminosities of
the \rk{E template}{typical ellipticals}. The Sd model neither matches the SED
of the E galaxy nor the SED of an S0 galaxy in terms of luminosities.  However,
considering galaxy mergers and the involved increase in mass by a factor of
$\approx 2$ for equal type mergers, all spiral model SEDs are able to match the
overall SED luminosities of the S0 galaxies approximately 4 Gyr after their
starbursts. This is in good agreement with results obtained by
\citet{Bicker2003Ap&SS}.

The lower part of Fig. \ref{sed_S0_Sa_10} shows the magnitude difference between
the Sa burst model and the S0- and E-type model. We removed the offsets due to
different stellar masses presented above by matching the I-band magnitudes of
all models. This emphasizes the extremly similar shapes between the post-burst
Sa model and both E and S0. Colours longwards of R are essentially
indistinguishable, only in the U-band are small differences. The E-type galaxy
has even redder U-B colours than the Sa burst model in its postburst phase,
while the \rk{S0 template}{S0-type model} is slightly bluer in U-B.

\begin{figure}
\includegraphics[width=\columnwidth]{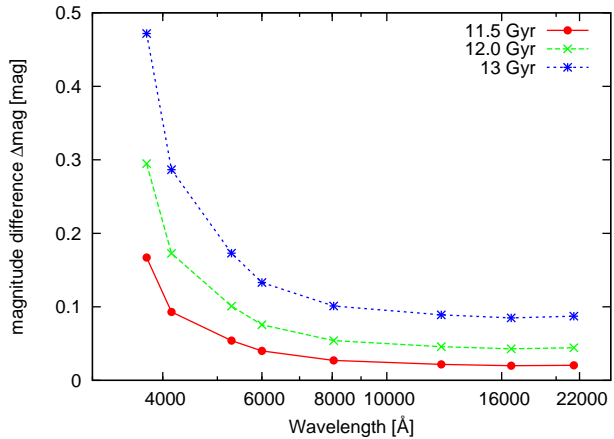}
\caption{Difference in magnitudes between the SEDs
  of an undisturbed and a burst Sa model at three different ages 11.5, 12 and
  13 Gyr. The undisturbed model was subtracted from the burst model (Sa, burst
  30\% at 11 Gyr, decline time 1.0 Gyr).}
\label{min_undist}
\end{figure}

In Fig. \ref{min_undist} we subtracted the SED of the undisturbed Sa model
from the SED of an Sa burst model with a 30\% burst after 11 Gyr and a decline
time of 1.0 Gyr at three different ages. This model shows one of the weakest
bursts of the grid and does not even have strong Balmer absorption lines.  We
only show the SEDs during and close to the E+A phase, i.e. at 11.5, 12 and 13
Gyr. Colours in the I through K-bands hardly change, only the luminosity
changes towards brighter magnitudes as a result of the additional built-up of
stellar mass during the burst. Towards shorter wavelengths the differences
increase with both increasing age and decreasing wavelength, reaching up to
$0.4$ mag in the U-band 2 Gyr after the burst. Since this model is one of the
models with the weakest burst of our grid, the differences given for this Sa
model are a lower limit. The differences for other models are larger and, for
example, amount to up to 1 mag for the Sd model with the maximum peak value
for H$\delta$. We therefore conclude that post-starburst galaxies, even after
very small bursts, can be distinguished from undisturbed galaxies on the basis
of their SEDs.

\begin{figure}
\includegraphics[width=\columnwidth]{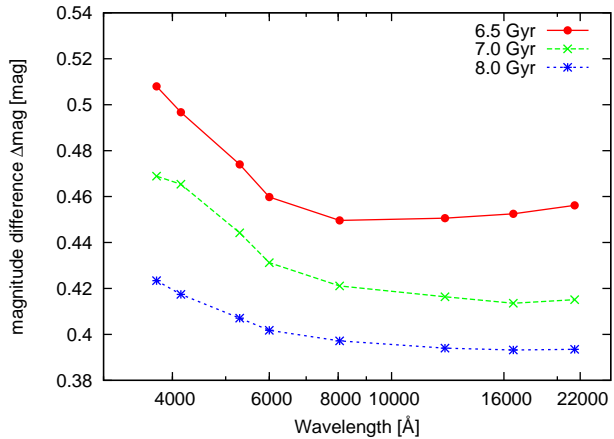}
\caption{Difference in magnitudes between the SEDs
  of two burst models at three different ages 6.5, 7 and 8 Gyr. The Sd model
  (burst 50\% at 6 Gyr, decline time 0.1 Gyr) was subtracted from the Sd model
  (burst 30\% at 6 Gyr, decline time 0.1 Gyr).}
\label{min_burst}
\end{figure}

In Fig. \ref{min_burst} we compare SEDs of two different burst models, both
starting as Sd-type model and both encountering bursts with decline time $0.1$
Gyr after 6 Gyr, but with different burststrengths of 30\% and 50\%.

We subtracted the SED of the weaker burst from the SED of the stronger burst,
so that all values in Fig. \ref{min_burst} translate directly into magnitude
differences between the two bursts in the sense that positive value mean that
the stronger burst is brighter.  The SEDs for ages of 6.5, 7, and 8 Gyr are
shown.  Although the stronger burst is brighter on average, there are only
very small differences between their colours; identical colours in this
representation would result in horizontal lines.

At 6.5 Gyr the maximum difference in colours is 0.05 mag and at 8 Gyr even
less. Obviously some very similar models can not be clearly distinguished from
each other by colours alone.  However, the difference in luminosities is 0.4
to almost 0.5 mag, which makes it possible to distinguish between models by
luminosity.  

We therefore conclude that it is possible to distinguish E+A galaxies,
post-starburst galaxies in a phase with still measurable {\oii} lines, and even
post-starburst galaxies with weak H$\delta$-lines from undisturbed galaxies and,
to a certain limit, from each other by carefully comparing their UV/U-opt-NIR
SEDs.

The advantage of this SED-based method is that only photometric observations
have to be obtained while no spectroscopic line measurements are necessary.
Post-starburst galaxies with emission lines, which would be excluded by the
classical E+A definition, can be detected and interpreted with this method, as
well as post-starburst galaxies for which the burst was not strong enough to
produce strong H$\delta$-lines with EW(H$\delta$) $\geq$ 5 \AA. It is also
possible to distinguish between most of the models, i.e. we can identify the
most likely progenitor galaxy type and estimate the burst strength.  We can thus
investigate the full range of post-starburst galaxies without the selection
effect we have identified in \citetalias{FalkenbergFritze09}, which excludes a
significant number of post-starburst galaxies from the E+A class.

The U, B and V bands are the most important ones and hence should all be
observed, while it is sufficient to have only one band in the near-infrared
(NIR).  The spectra in \citetalias{FalkenbergFritze09} indicate that SEDs with additional
filters in the UV, e.g. the FUV and/or NUV band from GALEX, would reveal even
more significant differences between the SEDs of post-starburst galaxies.


\section{Comparison to MORPHS data}\label{chapter_obs}
We will now use our models to study observed E+A galaxies. We present
the observational data and investigate them in terms of the Lick index
H$\delta$, luminosities, and colours. 
We also show how the best models can be found by 
comparing the SEDs of models and observations.

We adopt a cosmological model with $\rm H_0=73\,km\,s^{-1}\,Mpc^{-1}$, $\Omega_M=0.27$,
$\Omega_{\Lambda}=0.73$, and $\rm z_{form}=10$ \citep{Spergel+07}. The last parameter, the
formation redshift $\rm z_{form}$ is the assumed redshift of galaxy formation
$z_\mathrm{form}$. We need this parameter to convert redshifts into galaxy ages
at this redshift

\begin{equation}
\rm T_{gal}(z)=T(z)-T(z_{form})
\end{equation}

where $\rm T(z)$ is the Hubble time, i.e. the age of the universe, at redshift
z. Tab. \ref{tgal} shows the galaxy age $\rm T_{gal}$ as a function of
redshift for the cosmology we chose. Galaxy ages and redshifts used in our
models below are shown in bold font.

\begin{table}
  \caption{Redshift as a function of galaxy age $\rm T_{gal}$ for $\rm
    H_{0}=73\,km\,s^{-1}\,Mpc^{-1}$, $\rm\Omega_{M}=0.27$, $\Omega_{\Lambda}=0.73$, and $\rm
    z_{form}=10$.}\label{tgal} 
\centering

\begin{tabular}{l l }\hline \hline
$T_\mathrm{gal}$ [Gyr] & Redshift $z$ \\ \hline 
$1$            & $4.14$\\
$2$            & $2.62$\\
\textbf{3}   & \textbf{1.87}\\
$4$            & $1.40$\\
$5$            & $1.08$\\
\textbf{6}   & \textbf{0.84}\\
$7$            & $0.65$\\
\textbf{7.5} & \textbf{0.57} \\
$8$            & $0.49$\\
\textbf{8.4} & \textbf{0.44} \\
\textbf{9}   & \textbf{0.36}\\
$10$           & $0.25$\\
\textbf{11}  & \textbf{0.15}\\
$12$           & $0.065$\\ \hline
\end{tabular}

\end{table}

\subsection{The MORPHS E+A catalog}\label{catalog}
For the subsequent analysis we use data from the MORPHS
collaboration\footnote{The data and catalogs can be found online at
  http://astro.dur.ac.uk/\textasciitilde irs/morphs} \citep{Poggianti1999ApJ,
  Dressler1999, Oemler1999ASPC}. They used the Wide Field Planetary Camera 2
(WFPC2) onboard the Hubble Space Telescope (HST) to study the morphology of
galaxies in ten distant clusters in a redshift range of z=0.37-0.56
\citep{Smail1997ApJS}.  

The catalog lists redshift z, quality of the spectra, EW(H$\delta$),
luminosities, and colours for 88 possible E+A galaxies.
The necessary spectroscopic data were obtained with the 200 inch (5.1 m) Palomar
Hale Telescope (P200), the 4.2m William Herschel Telescope (WHT), and the 3.5m
New Technology Telescope (NTT). Photometric data were taken with the 4-Shooter
cameras on the 200 inch Palomar Hale Telescope and with HST using WFPC2.
Although the data have been taken with various telescopes, they were processed
with the same data reduction methods to ensure homogeneity.

For most galaxies r, g$-$r and i in the Thuan \& Gunn system \citep[similar
    to the SDSS filters, magnitudes are in AB system]{Thuan1976PASP} of the
P200 are given. Data from HST in I (F814W), R (F702W), B$-$I (F450W$-$F814W) and
V$-$I (F555W$-$F814W) of the HST filter system are given for only a few
galaxies. To avoid uncertainties related to transformations between filter
systems we therefore calculated apparent magnitudes in all those filters by
computing spectra first which are then convolved with the appropriate filter
response curves.

\subsection{Sample selection}
Galaxies with EW(H$\delta$) $\geq$ 3 \AA\ and no significant emission lines are
classified as E+A galaxies. There is a subdivision into H$\delta$-stronger E+A
galaxies with H$\delta$ $\geq$ 8 \AA, called a+k galaxies, and H$\delta$-weaker
galaxies, named k+a, with $\rm 5\,\AA\le\ewhd\le8\,\AA$.  We excluded all
galaxies with doubtful classification as well as spectra of quality 4 in the
MORPHS grading scheme \citep[where 1 is best and 5 is worst]{Dressler1999} to
ensure reliable measurements of EW(H$\delta$). We also exclude galaxies with
EW(H$\delta$) $<$ 5 \AA, because H$\delta$-values below this limit can easily be
reached by non-poststarburst galaxies as we shown in
\citetalias{FalkenbergFritze09}.  Applying these selection criteria leaves us
with a final sample of 44 E+A galaxies.

\subsection{Lick Index H$\delta$, Luminosities and Colours of Observed E+A
  Galaxies}
\label{obs_data}
\begin{figure}
\includegraphics[width=\columnwidth]{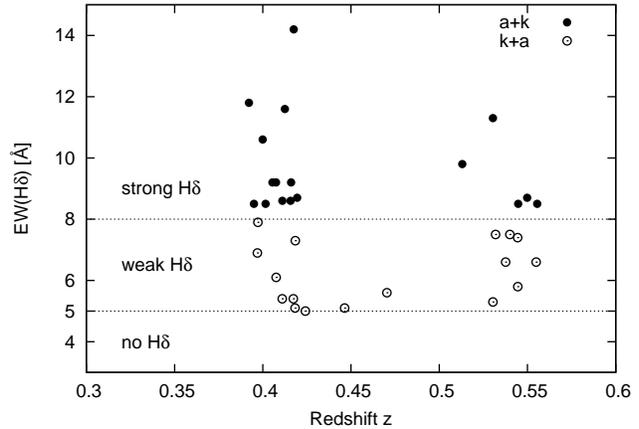}
\caption{EW(H$\delta$) versus redshift z of E+A galaxies observed by the
  MORPHS collaboration.}
\label{Obs_Hd_z}
\end{figure}

In Fig. \ref{Obs_Hd_z}, the EW(H$\delta$) of all galaxies from the catalog are
shown versus redshift. Filled circles indicate H$\delta$-strong a+k galaxies,
open circles show H$\delta$-weaker k+a galaxies. It can be seen that there is a
group of galaxies at redshift z$\simeq$0.4 and another group at $z\simeq 0.55$
corresponding to the redshifts of clusters studied by the MORPHS
collaboration. The values for EW(H$\delta$) range from 5~\AA\ (our selection
  criterion) to 14.2~\AA. \rk{However, the value of 14.2 \AA\ is marked as
  questionable in the catalog. The maximum secure value for EW(H$\delta$) is
  11.3 \AA.}{} \rk{Note also that 5~\AA\ is the selection criterion for the E+A
  class chosen by us.}{} The measurements of EW(H$\delta$) are reproducible to
$\pm$10\% for H$\delta \geq$ 5 \AA\ according to \citet{Dressler1999}.

\begin{figure}
\includegraphics[width=\columnwidth]{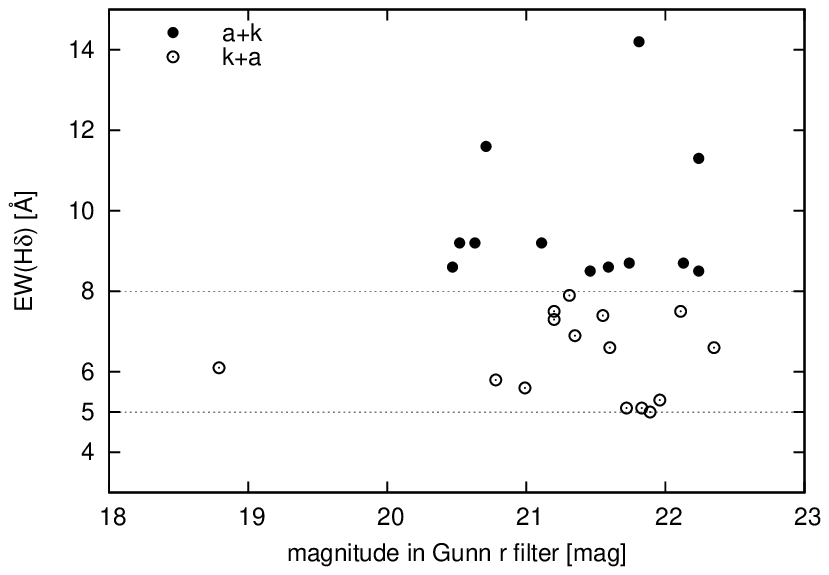}
\includegraphics[width=\columnwidth]{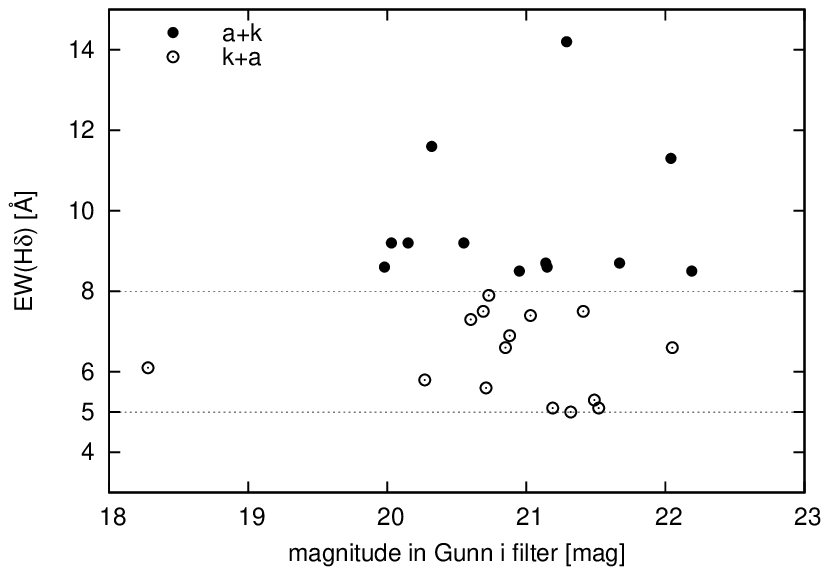}
\caption{EW(H$\delta$) versus Gunn \& Thuan r and i luminosities of E+A
  galaxies observed by the MORPHS collaboration.}
\label{Obs_Hd_r}
\end{figure}

In Fig. \ref{Obs_Hd_r}, the EW(H$\delta$) is plotted versus r- and i-band
luminosities of the Thuan \& Gunn system. Filled and open circles again indicate
a+k and k+a galaxies, respectively.  The systematic errors for magnitudes and
colours are estimated by \citet{Dressler1992ApJS} to be of the order of
$0.1-0.2$ mag. These observed values of EW(H$\delta$) are to be compared
  with the evolution of our different model EW(H$\delta$) during their
  post-burst phases as shown in \citetalias{FalkenbergFritze09}, Figs. 4-6. We
  recall that the maximum EW(H$\delta$) reached by our models was around 8.5
  \AA. Please note that the two galaxies with the highest values for
EW(H$\delta$), located in the upper right part of the Fig. \ref{Obs_Hd_r}, are
marked as questionable.  We notice that there is an outlier in luminosity at r-
and i-band luminosities of 18.25 mag and 18.28 mag, respectively. However, there
is no note in the catalog that this galaxy has questionable values. If the
luminosities and the E+A classification for this galaxy are correct, the
brightness of this galaxy can only be explained by a strong previous burst in a
merger of two previously very bright galaxies.

\begin{figure}
\includegraphics[width=\columnwidth]{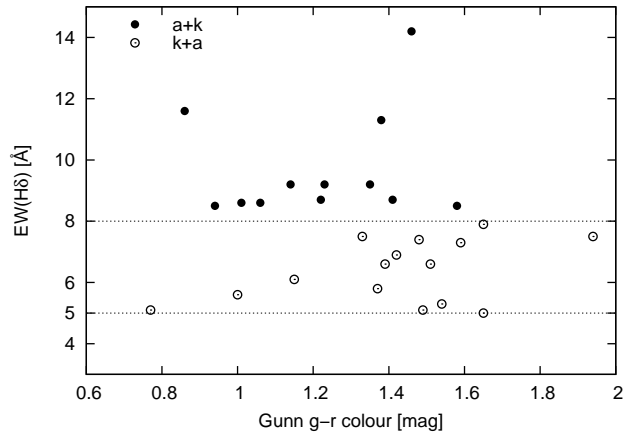}
\caption{EW(H$\delta$) versus colour for 
Gunn \& Thuan g$-$r of E+A galaxies observed by the MORPHS collaboration.}
\label{Obs_Hd_V-I}
\end{figure}

Fig. \ref{Obs_Hd_V-I} shows the H$\delta$ equivalent width EW(H$\delta$) as
function of the g-r colour in the Thuan \& Gunn system.  From the g-r colour
plot we find that k+a galaxies are redder than a+k galaxies showing higher
EW(H$\delta$). This is in agreement with our models which show that galaxies
with high EW(H$\delta$) can only have blue colours while galaxies with low
EW(H$\delta$) can be either blue or red \citepalias[see][]{FalkenbergFritze09}.

The provided sample with HST I and R luminosities as well as B-I and V-I
colours is too small to make any statistical statements. The individual
galaxies can still be analyzed, however, this is not the subject of this
paper.

\subsection{Methods of the Comparison between Observations and Models}
To compare our models with the observational data from the MORPHS catalog, we
first have to ensure that EW(H$\delta$), colours and magnitudes are all given on
the same system. For this purpose we implemented the Thuan \& Gunn filter
system of the Palomar Hale Telescope into the GALEV code and convolved the
spectra of our model grid with the Thuan \& Gunn filter system and with the HST
filter system.

Our GALEV models currently include the H$\delta$ definition of \citet[see
  \citetalias{FalkenbergFritze09} for details]{Trager1998ApJS} which is slightly
different from the definition of \citet{Balogh1999} used by the MORPHS group. To
ensure that this does not affect any of our results we measured EW(H$\delta$) of
an A-star spectrum degraded to the resolution of the MORPHS spectra using both
definitions.  The result is that the maximum deviation amounts to 0.2 \AA, while
the measurement uncertainties for the EW(H$\delta$) is given by
\citet{Dressler1999} to be of the order of 10\%, which corresponds to an
uncertainty of 0.5 \AA\ for an EW(H$\delta$) of 5 \AA\ (the lowest value for
EW(H$\delta$) of E+A galaxies).  We therefore conclude that the EW(H$\delta$) as
calculated by GALEV and the measured EW(H$\delta$) of the MORPHS group are
equivalent within the measurement uncertainties.

\cite{Dressler1999} and \cite{Dressler2004} furthermore compared the measurement
of $\ewhd$ using the bandpass definition and direct line profile fits, the
latter of which is used by the MORPHS group. They found that for weak but
measurable cases, i.e. $\ewhd\approx 5\, \AA$, the line-fitting and the bandpass
methods are equivalent.  For strong absorption lines, however, the bandpass
definition tends to underestimate the EW, because in these cases \rk{the
  pseudo-continuum is already depressed relative to the true continuum, leading
  to lower equivalent widths.}{the large population of early A-type stars with
  relatively broad H$\delta$ that is necessary to reach these line strengths
  leads to a depression in the continuum bands, which in turns lowers the
  measured line strengths. This issue of measuring the line strength of
  H$\delta$ in an optimal way is discussed in length in \cite{Dressler2004}.}
This explains why the observations in Figs. \ref{Obs_Hd_z}-\ref{Obs_Hd_V-I}
reach EW(H$\delta$) as high as $14$ \AA, while the maximum value found in our
model grid is only around $8.5$ \AA. This does, however, in no way affect any
of our results or conclusions.

\subsection{Results from the Comparison between Observations and Models}
\label{compare}

As we found in \citetalias{FalkenbergFritze09}, \rk{the classical definition of
  E+A galaxies used by the MORPHS group excludes all post-starburst galaxies
  with a SFR decline time $\tau \geq$ 0.3 Gyr}{we can not reproduce the MORPHS
  observations by any of our models with SFR decline time $\tau \geq$ 0.3 Gyr:
  Models with longer decline times either show remaining SF during the H$\delta$
  strong phases or are no longer H$\delta$ strong as soon as the SF has
  decreased sufficiently}. However, we will not exclude the models with a longer
decline time to see how good the comparison of luminosities and colours between
models and observations is when trying to find matching models.

To compare MORPHS data at different redshifts with our models, the model spectra
have to include cosmological and evolutionary effects as described in
Sect. \ref{sect:galev}. \rk{}{Evolutionary (e-) corrections describe the fact
  that galaxies at redshift $z=0.4$ are $\approx 4$ Gyr younger than local galaxies
  at $z=0$. Cosmological (k-) corrections on the other hand describe the
  bandpass shift due to cosmic expansion: The observed r-band magnitude (central
  wavelength: $\rm\lambda_C \approx 6700$ \AA) resembles approximately the
  rest-frame g-band ($\rm\lambda_C = 5100 \AA \approx 6700\,\AA / (1+0.4)$). To
  prevent further uncertainties from entering analysis we decided to not change
  the observed data, but rather include both e- and k-corrections into our
  models}.

At first we investigated which galaxy models are in their E+A phase in
the redshift range between $z=0.36-0.56$.  Most models can already be excluded
before any comparison with the data, because they are not in their
H$\delta$-strong phase in the redshift range of interest. Of course, all
galaxies with an onset of burst or truncation/termination at $z\simeq0.35$ can
be excluded, because they are still undisturbed at $0.36\leq z \leq 0.56$.

The galaxy models with an onset at $z\simeq 0.93$ can also be excluded because
they have already passed through the H$\delta$-strong phase.  ALL post-starburst
galaxies with an onset at $z\simeq 0.44$ reaching EW(H$\delta$) $\geq$ 5
\AA\ are possible. However, the galaxies with an onset at $z=0.44$ only cover
the low redshift part of the redshift range during their H$\delta$-strong
phase. We can therefore already predict that galaxy models with an onset between
$z=0.44$ and $z=0.93$ would fit the data much better.  As examples a few galaxy
models with an onset at $z=0.57$ were calculated in addition to the grid
explored so far.

\rk{}{We furthermore chose to not only use colours but also absolute magnitudes
  in the subsequent analysis, hence not allowing galaxy masses as completely
  free parameter. The physical reason for this is that the properties during and
  after the starburst heavily depend not only on the amount of newly born stars,
  but also on the star formation history before the burst (see
  Sect. \ref{chapter_sed}). This spectral type dependence then translates into a
  mass-dependence, since galaxies of each type only cover a narrow mass-range as
  derived from absolute magnitudes in the local universe
  \citep[e.g.][]{Sandage1985AJa, Sandage1985AJb}. This limitation hence rules
  out a number of otherwise possible scenarios, e.g. the formation of bright
  E+As by pure truncation of SF.}

\begin{figure}
\includegraphics[width=\columnwidth]{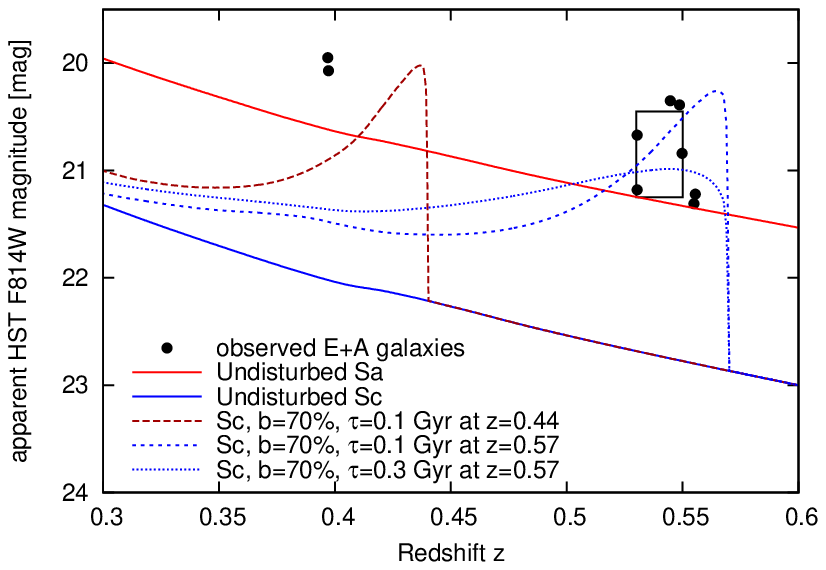}
\includegraphics[width=\columnwidth]{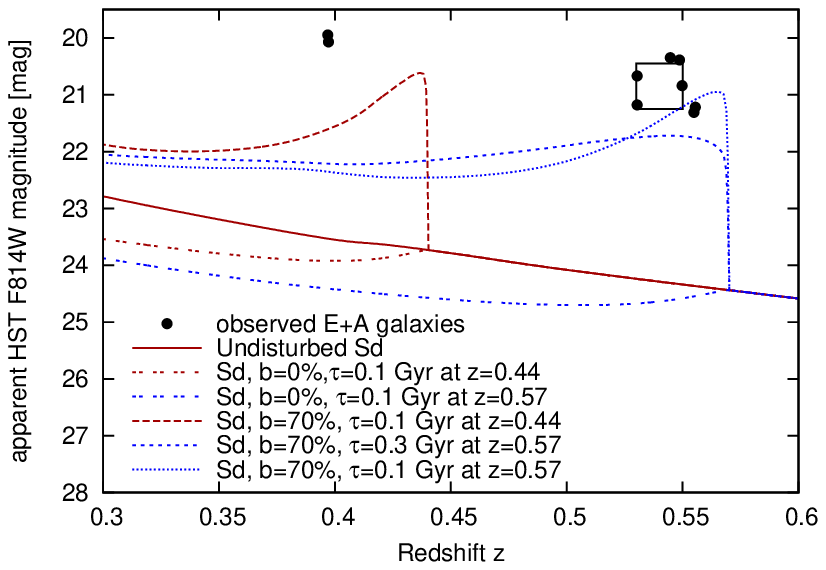}
\caption{HST F814W I-band luminosity versus redshift z. The symbols show the
  observed data. The box indicates the mean values with
  1 $\sigma$ standard deviations of the observation
  data. The lines show some of the models with onset at $z\simeq 0.44$ and
  $\simeq 0.93$ (top) and $z\simeq 0.57$ (bottom).}
\label{I_z}
\end{figure}

\begin{figure}
\includegraphics[width=\columnwidth]{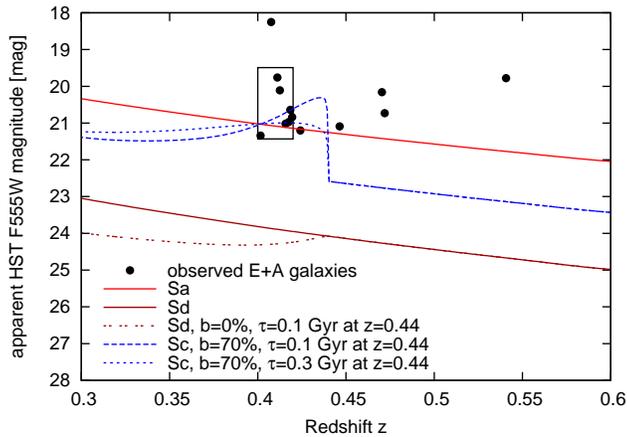}
\caption{HST F555W V-band luminosity versus redshift z. The symbols show the
  observed data. The box indicates the mean values with 1 $\sigma$ standard
  deviations. The lines show some of the models with onset at $z\simeq 0.44$.}
\label{R_z}
\end{figure}

\begin{figure}
\includegraphics[width=\columnwidth]{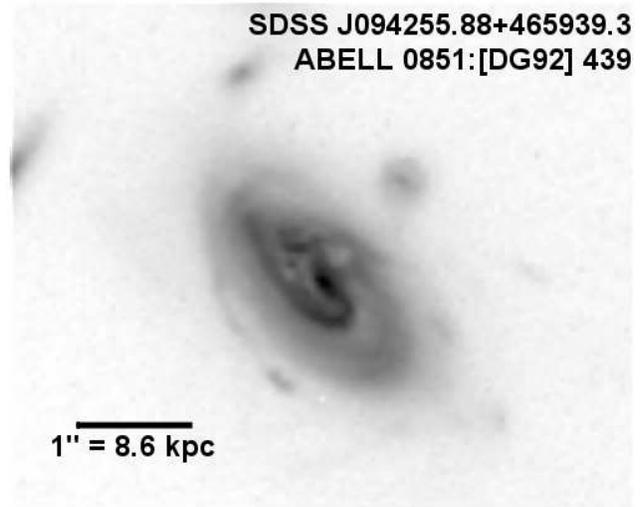}
\caption{R-band image of CL0939+4713\#18, taken with WFPC2. This galaxy shows a
  very irregular appearance with multiple (tidal) tails sticking out of the main
  body.}
\label{fig:groupmerger}
\end{figure}

\begin{figure}
\includegraphics[width=\columnwidth]{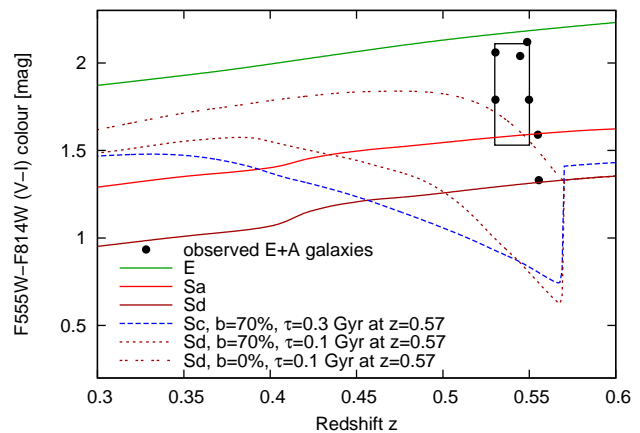}
\caption{HST F555W-F814W colour ($\sim$ V$-$I) versus redshift z. The symbols
  show the observed data. The box indicates the mean values with 1 $\sigma$
  standard deviations. The lines show some of the models with onset at $z\simeq
  0.57$.}
\label{V-I_z}
\end{figure}

\begin{figure}
\includegraphics[width=\columnwidth]{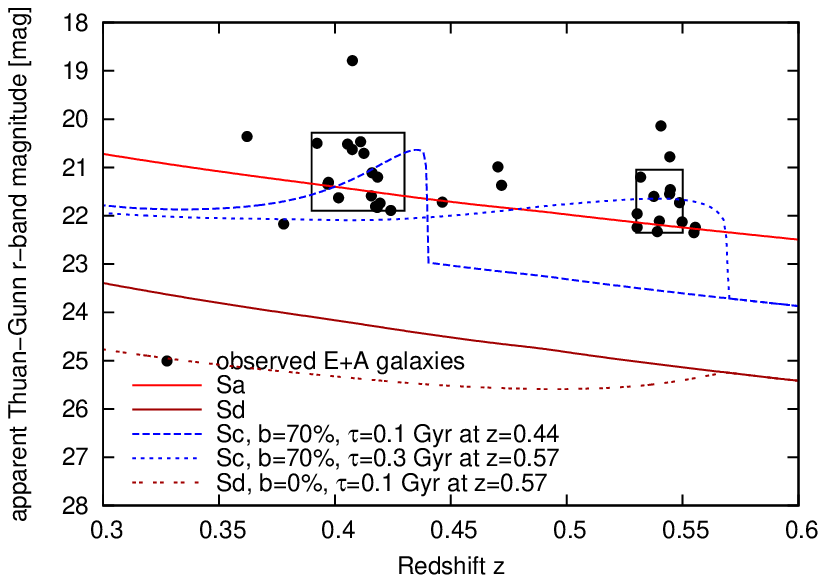}
\includegraphics[width=\columnwidth]{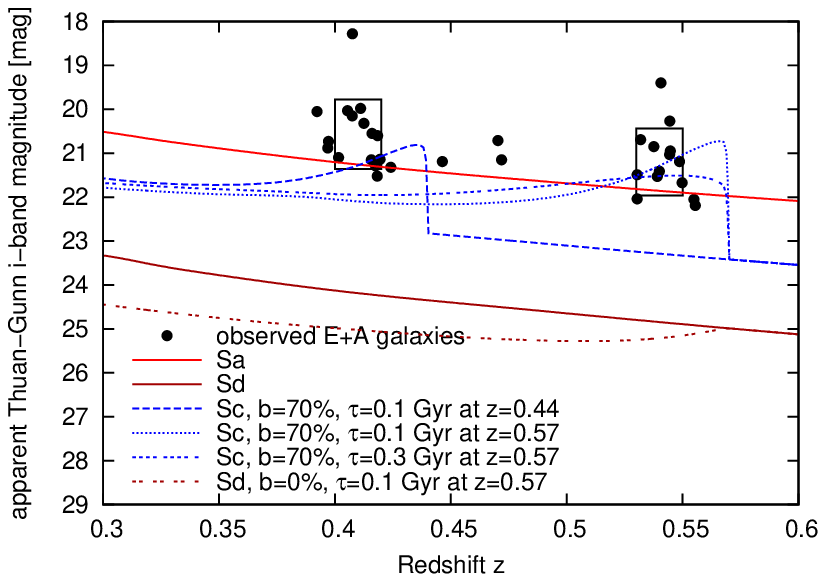}
\caption{Thuan \& Gunn luminosity in the r- and i-band versus redshift z. The
  symbols show the observation data. The box indicates the mean values with 1
  $\sigma$ standard deviations of the observation data. The lines show some of
  the models at $z\simeq 0.44$ and $z\simeq 0.57$.}
\label{r_z}
\end{figure}

\begin{figure}
\includegraphics[width=\columnwidth]{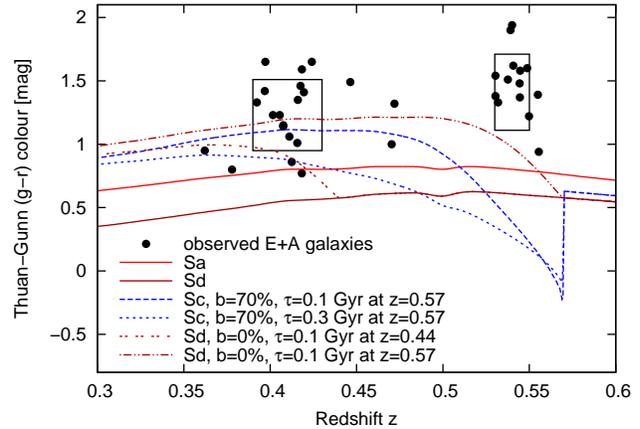}
\caption{Thuan \& Gunn colour (g$-$r) versus redshift z. The symbols show the
  observation data. The box indicates the mean values with 1 $\sigma$ standard
  deviations of the observation data. The lines show some of the models at
  $z\simeq 0.44$ and $z\simeq 0.57$.}
\label{g-r_z}
\end{figure}

In Fig. \ref{I_z}-\ref{g-r_z}, we present luminosities and colours in various
filters as function of redshift. We recall that our undisturbed model SFHs
  were chosen as to match after 13 Gyr of undisturbed evolution the observed
  galaxies' average colours. Model masses were chosen as to yield the average
  B-band luminosities observed for Sa, Sb, Sc and Sd galaxies (see
  \citetalias{FalkenbergFritze09}, Sect. 2.1). Observed undisturbed local
  galaxies show 1 $\sigma$ ranges around those mean luminosities that need to be
  considered in the comparison between observations and models.
Very important to note is that on all the plots of EW(H$\delta$),
luminosities, and colours as a function of redshift the models evolve from
the right-hand high redshift side towards the left-hand low redshift side in
contrast to all the time evolution plots shown in \citetalias{FalkenbergFritze09}.

Boxes indicate the mean values with 1 $\sigma$ ranges of the observational data
in our sample. Shown are those models that have the maximum
and minimum values of the grid and have their H$\delta$-strong phase in the
given redshift range.  We also show the undisturbed Sa and Sd models for
comparison.

At the top of Fig. \ref{I_z}, WFPC2 I-band luminosity is plotted versus
redshift.  The Sc galaxy with a burst of 70\% and a decline time of 0.1 Gyr at
$\simeq 0.44$ has a luminosity, that matches quite well the two observed E+A
galaxies at a redshift of $z \simeq 0.4$ (all models with onset $z\simeq 0.44$
do so). Remember that the luminosities given in all plots are representing the
average luminosity for each particular galaxy type, while true galaxies span a
range in luminosities (see Tab. \ref{tab:sigmaranges}). Furthermore in the case
of galaxies with bursts, those are induced by merging two galaxies, increasing
the effective mass and hence luminosity by a factor of 2, corresponding to a
magnitude difference of $0.75$ mag. Magnitudes of E+A galaxies at $\rm z\approx
0.5$ can be reproduced by earlier bursts. We show two examples of galaxies
starting as Sc galaxies and encountering strong bursts of 70\% with decline
times of 0.1 and 0.3 Gyr. Both very well match the average i-band magnitudes
observed at $\rm z\approx 0.5$.

We also show for comparison the magnitude evolution of the undisturbed Sa and Sd
models. The Sd is too faint by several magnitudes. The Sa-type model, however,
is found to have approximately the right luminosity to match the
observations. However, as we have shown in \citetalias{FalkenbergFritze09},
Sa-type galaxies do not become H$\delta$ strong enough to be considered E+A
galaxies, since they are lacking the gas reservoir to form a sufficient amount
of new stars. Minor mergers of one early-type spiral and one late-type spiral 
could be a way to increase the gas-reservoir and turn early-type spirals into
E+As. The detailed modelling of the range of possible unequal-type mergers,
however, is beyond the scope of the present paper.

In the bottom panel of Fig. \ref{I_z} we compare bursts occuring in previously
undisturbed Sd-type galaxies. Bursts start either at $\rm z=0.57$ or $z=0.44$
and convert 70\% of the then available gas mass into stars; decline times shown
are 0.1 and 0.3 Gyr. We manage to reproduce the luminosities of all E+A galaxies
if we also account for the doubling of mass during the merger. Truncation
scenarios, here shown for $\tau=0.1$ Gyr and two different onset redshifts, however, are significantly too faint to match the observations.

In Fig. \ref{R_z} we show the HST F555W ($\sim$ V) band magnitudes of all
galaxies observed in this filter. Again we find that bursts of Sc galaxies with
short decline times match the observations. Galaxies at higher redshifts can be
explained by slightly earlier bursts with onset redshifts of $0.5\ldots
0.7$. The only exception is one extremely bright galaxy with $m_{R} \approx
18.2$ mag or $\rm M_{R}=-23$ mag using a distance modulus of $\rm m-M=41.4$
(Galaxy \#18 of cluster CL0939+4713 in the catalog of \cite{Dressler1999} and \#
224 in catalog of \cite{Smail1997ApJS}). This can only be explained by a
massive merger of two above-average massive galaxies, or a complex merger of
more than two galaxies, possibly a small, infalling group. Further evidence for
this scenario is the complex shape (see Fig. \ref{fig:groupmerger}) showing
significant substructure with multiple (tidal) tails as well as several nearby
companion galaxies. However, on the basis of currently available data we are not
able to make more definite statements on the nature of this galaxy.

Fig. \ref{V-I_z} shows the HST F555W$-$F814W ($\sim$ V$-$I) colour. We only show
some examples for scenarios with burst beginning at redshift $z\simeq 0.57$,
because all observed galaxies lie at $z\simeq 0.55$. The majority of the data
points can be explained with these models. However, for some observed galaxies
with red V$-$I colours, an even earlier onset at $\rm z\geq 0.6$ has to be
considered.

In the top panel of Fig. \ref{r_z} the r-band luminosity in the Thuan \& Gunn
system is plotted versus redshift. All galaxies at z $\approx0.4$ and z
$\approx0.5$ can be described by our models with strong bursts and decline times
of 0.1 Gyr and 0.3 Gyr. The truncated Sd model is found to be too faint, in
agreement with our findings for the HST filters above.

The bottom panel of Fig. \ref{r_z} shows the Thuan-Gunn i-band magnitude of
all E+A galaxies. As found already for the g-band magnitudes we can reproduce
the full spectrum of observed magnitudes. An important finding is that
galaxies which encounter a burst at redshift z $\approx 0.57$ are still bright
enough to fit the data for those E+As at z $\approx 0.4$ when we account for
the 0.75 mag increase due to the doubling of mass in the merger. 

In both panels of Fig. \ref{r_z} we also show the evolution of undisturbed Sa
and Sd-type galaxies. Although the Sa-type model reproduces the required
luminosities, both are not in an E+A phase at their particular age. Likewise
in both plots we have one significant outlier at too bright magnitudes. This
is the galaxy we have already discussed above.

Summarising our discussion of model magnitudes as compared to observations we
can conclude that many models can already be excluded on basis of magnitudes
in individual filters alone. Burst models in general reach the observed
bright magnitudes, while models with SF truncation without previous starburst
are too faint to match the observations.

Fig. \ref{g-r_z} shows the g$-$r colour versus redshift. During and shortly
after the burst our models are too blue to match the observations. However,
models with strong bursts occuring at redshift z $\approx 0.57$ reach the red
colours observed in E+A galaxies at redshift z $\approx 0.44$. This is a very
important result, since the time between those redshifts accounts to $\simeq1$
Gyr ($\rm t_{gal}(z=0.57)=7.3$ Gyr and $\rm t_{gal}(z=0.44)=8.2$ Gyr). After
this time the SFR has declined sufficiently to not show up in the [O II] lines
anymore while EW(H$\delta$) reaches a maximum. We can hence fully reproduce
all four factors determining E+A galaxies at those redshifts: 1) strong
absorption in H$\delta$; 2) no emission in [O II]; 3) red colours 4) high
luminosities.

Galaxies observed in their E+A phase at redshift z $\simeq 0.5$ can be
described by earlier bursts occuring at $z\approx 0.75$.

Furthermore, only models with short decline times of $\le~0.3$ Gyr match the
luminosities and colours, confirming our results based on 
the EW(\oii) described earlier. It seems that the
constraints from luminosities and colours provide very valuable complementary
information to the studies of absorption lines and allow to discriminate
between possible and impossible scenarios. This proves that the comparison
of observed and model SEDs introduced in Sect. \ref{chapter_sed} is a good
method to find adequate models for the observations.

\subsection{Comparison of Observed SEDs with Model SEDs}
\label{sed}
\begin{figure}
\includegraphics[width=\columnwidth]{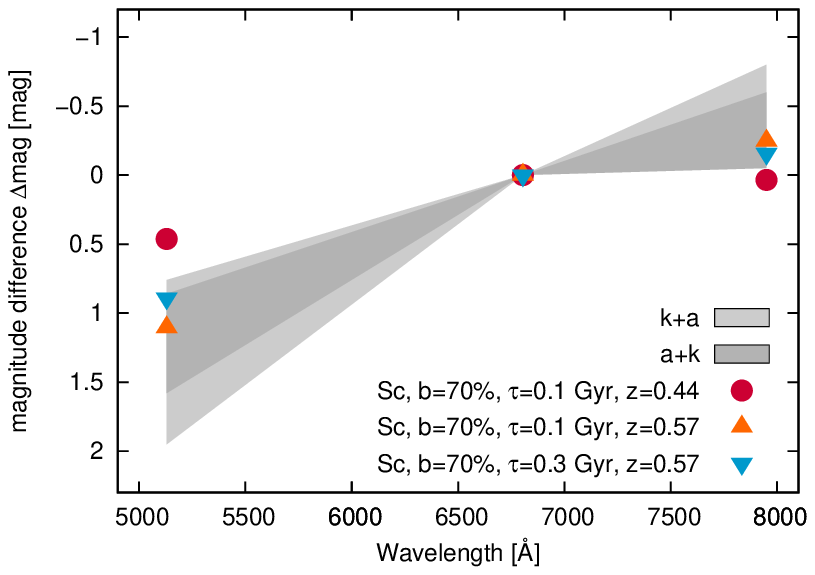}
\includegraphics[width=\columnwidth]{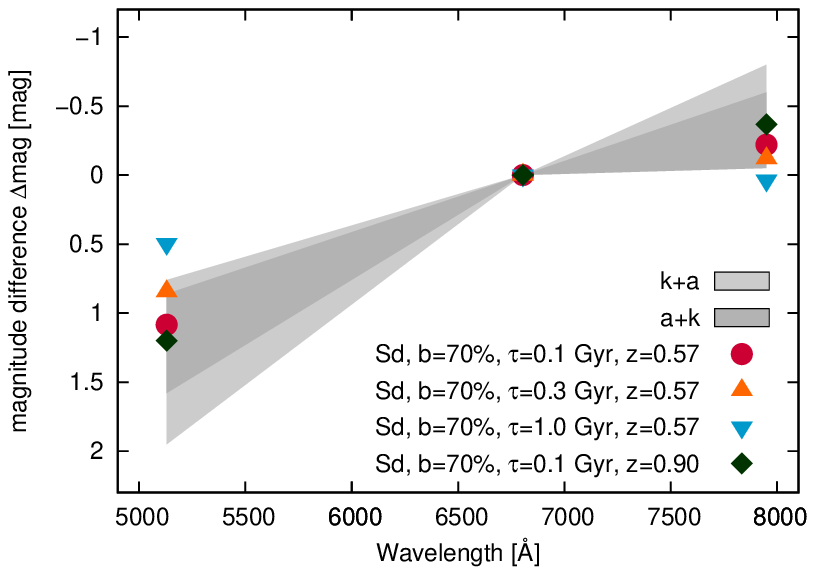}
\caption{Comparison of full SEDs. The range spaned by the observed a+k (k+a)
  galaxies are shown as dark (light) shaded grey region, In the top panel we
  show SEDs for three burst models with Sc progenitors, in the lower panel we
  show four models with Sd-type progenitor galaxies.}
\label{fig:easeds}
\end{figure}

In the following we will compare the full model SEDs to the observed SEDs.
Since we have found earlier that our model magnitudes match the average
observed magnitudes for the E+A galaxies, we normalize all SEDs to a r-band
magnitude of 0, so that our SEDs essentially consist of the g-r and i-r
colours. We do not show the HST magnitudes, since the majority of galaxies in
the MORPHS catalog were only measured in the Thuan-Gunn system. Fig.
\ref{fig:easeds} shows the observed colour range of all a+k and k+a galaxies
as grey-shaded areas. In the top panel of this Fig. we also show three
different model-SEDs based on Sc-type progenitor galaxies for redshift
$z=0.4$. In accordance with our analysis of the individual colours we find
that the model with burst at $z=0.44$ is still too blue to explain the
observations. Models with earlier bursts at $z=0.57$ lie within the observed
colour range, with the shorter burst-duration of $0.1$ Gyr providing a better
match than the model with the slightly longer duration of $0.3$ Gyr.

The lower panel of Fig. \ref{fig:easeds} compares the observed data with
Sd-progenitor models. Both models with burst redshift of $\rm z=0.57$ and burst
durations $\le 0.3$ Gyr lie within the observed range. The model with
the longer burst duration of 1 Gyr is still too blue at redshift $z=0.4$ for
which the SEDs are shown. For comparison we also show an SED of an earlier
burst at $\rm z=0.9$. Although this matches the observed SED quite well, the
galaxy model is no longer in an E+A phase at the observed redshift $z=0.4$, so
that this model can be excluded as well.

\section{Where does the remaining gas go?}
\label{sect:gasfate}
One question remains yet untouched: The fate of the remaining gas that was not
converted into stars during the burst. In the following we will show that the
energy input by supernovae is sufficient to remove this gas. Integration of an
Salpeter IMF with mass-limits of 0.1 and 100 $\msun$ yields that $\approx
10\%$ of the gas-mass is converted into stars with $\rm M>8 \msun$, that end
their life as core-collapse supernovae. We will base our example on a
burst-model starting as Sc-type model with total mass of
$1.4\times10^{10}\,\msun$ and encountering a strong (70 \%) burst at an age of
$7.5$ Gyr. This burst forms a total stellar mass of $4.8\times10^9\,\msun$. We
further assume a typical mass of $\rm 10\msun$ for stars with $\rm M>8\,\msun$
and that each resulting supernova produces a total energy of $\rm 10^{51}$ erg
\citep{Burrows+95,Hamuy03}. All supernovae formed during the burst hence
produce a energy output of $\approx 5\times10^{58}$ erg. To derive the
potential energy needed to remove the remaining $2\times 10^9\msun$ gas we use
the total mass of the galaxy and assume a typical radius of $\rm 3\,kpc$. This
result in a binding energy of $\approx 7.3\times 10^{57}$ erg, or 14\% of the
energy produced by the supernovae. This fraction is in good agreement with
earlier results of $0.09$ \citep{Thornton+98} and $0.1-0.2$ \citep{Cole+94}.

 
\section{Summary and Outlook}\label{conclusion}
Following the first part of our study presented in
\citetalias{FalkenbergFritze09} we present Spectral Energy Distributions for yet
undisturbed, pre-starburst galaxies, galaxies with ongoind starbursts and
post-starburst galaxies. We showed that comparing the SEDs of post-starburst
galaxies to those of undisturbed spiral, S0 and E galaxies as well as among each
other is a sufficient method to select and investigate post-starburst
galaxies. The interplay between the burst strength, the decline time, the onset
of the burst/halting of SF and the progenitor galaxy type shapes the SED in the
following way:

\begin{enumerate}
\item The later the \emph{progenitor galaxy type}, the more obvious is the
  difference between the SED in the post-starburst phase to the SED of the
  corresponding undisturbed galaxy model. Late-type spirals furthermore evolve
  more slowly towards S0 spectral types than early-type spirals.

\item The earlier the \emph{onset} of the burst the more luminous
  is the model SED in its starburst and post-starburst phases.

\item The higher the \emph{burst strength}, the more luminous is the model SED
  throughout its starburst and post-starburst phase.

\item The longer the \emph{decline time}, the longer takes the time evolution
  of luminosities and colours of the SED to evolve first into the SED of a
  post-starburst galaxy and later into the SED of an S0 galaxy. 

\end{enumerate}

We have shown that we are able to distinguish E+A galaxies, post-starburst
galaxies in an early phase after a burst with still measurable {\oii} lines, and
even post-starburst galaxies with weak H$\delta$-lines from undisturbed
galaxies and, to a certain limit, from each other by carefully comparing their
SEDs.

The advantage of this method is that only photometric observations have to be
made and no line measurements are necessary. Post-starburst galaxies with
emission lines, which are excluded by the classical E+A definition can be
detected and interpreted, as well as post-starburst galaxies for which the
burst was not strong enough to produce strong H$\delta$-lines with
EW(H$\delta$) $\geq$ 5 \AA. It is possible to distinguish between most of the
models i.e., we can identify the progenitor galaxy type and estimate the burst
strength. This means we can investigate the full range of post-starburst
galaxies without the selection effect which excludes a number of
post-starburst galaxies from the E+A class \citepalias[see][]{FalkenbergFritze09}. 

The most important bands crucial for this SED analysis are U, B and V, while
it is sufficient to have only one band in the NIR. Our model spectra in
\citetalias{FalkenbergFritze09} indicate that SEDs with additional filters in the UV, e.g. from GALEX, would
reveal even more subtle differences among the SEDs of different types of
post-starburst galaxies.

In the second part of this paper we compared our model results to the MORPHS
catalog containing 88 E+A galaxies in the redshift range z=0.36$-$0.56. We
constructed a grid of models spanning a wide range of progenitor galaxies,
burst redshifts, burst decline times and burst strengths. All our models fully
include both cosmological and evolutionary corrections as function of redshift
\textbf{and} galaxy type.  

We find that k+a galaxies are redder than a+k galaxies with higher
EW(H$\delta$). This is in good agreement with our models which show that
galaxies with high EW(H$\delta$) can only have blue colours while galaxies
with low EW(H$\delta$) can either be still blue or already red.

From our comparison of model and observational data, we find that only galaxies
with strong bursts $\approx 70\%$ and relatively short decline times $\le 300$
Myr can explain the observed colours and magnitudes.  This is in good agreement
with earlier studies \citep{Poggianti+96, Balogh+97, Balogh+05, Yan+08} which
also found indications for recent bursts forming a significant fraction of the
observed stellar masses. Weaker bursts do not reach the required H$\delta$
strengths, while models with SF truncation instead of a starburst are
significantly too faint, confirming earlier findings from
\cite{ChristleinZabludoff04} who found that simple fading without an increase of
the bulge luminosity does not reproduce observations of S0 galaxies.

We hence conclude that a large fraction of the parameter space can easily be
excluded based on SEDs alone. This is remarkable since our SEDs only consist
of three filters and only cover a short wavelength range of the spectrum. 

We could also show that supernovae being formed as a consequence of the burst
can produce enough energy to remove the remaining gas that was not turned into
stars from the galaxy, explaining why most observed E+A galaxies do not
contain significant amounts of cool gas.

The predictions from our models can be used to collect more and better data
for the SEDs of post-starburst galaxies.  With better sampled SEDs, in
particular with UV filters and one NIR band included, and a suitable SED
fitting tool (e.g. AnalySED \citep{Anders2004MNRAS} or GAZELLE (Kotulla \&
Fritze, in preparation)) it will become possible to improve the method outline
here and to better constrain not only progenitor galaxies and transformation
scenarios, but also limit the range of possible burst ages and burst
fractions.

Investigating SEDs of \emph{all} post-starburst galaxies promises to lead to a
better understanding of the transformation processes of galaxies in high
density environments.

\section*{Acknowledgments}
We thank our referee, Guy Worthey, for his patience and very insightful, helpful
and encouraging comments that greatly helped us to improve this paper.


\setlength{\bibhang}{3em}
\bibliographystyle{aa}
\bibliography{aylin2}

\label{lastpage}

\end{document}